\documentstyle[prd,aps]{revtex}

\begin{document}
\draft

\title{Classical Aspects of Accelerated Unruh-DeWitt Type Monopole
Detectors}
\author{Shih-Yuin Lin\footnote{
Electronic address: {\tt sylin@phys.sinica.edu.tw}}}
\address{Institute of Physics, Academia Sinica, Nankang, Taipei 11529,
TAIWAN }
\date{June 2001}
\maketitle

\begin{abstract}
We have shown the classical correspondence of Unruh effect in the classical
relativistic electron theory in our previous work\cite{lin}.
Here we demonstrate the analogy between the classical relativistic
electron theory and the classical Unruh-DeWitt type monopole detector
theory. The field configuration generated by a uniformly accelerated
detector is worked out. The classical correspondence of Unruh effect for
scalar fields is shown by calculating the modified energy density for the
scalar field around the detector. We conclude that a classical monopole
detector cannot find any evidence about its acceleration unless it has a
finite size.
\end{abstract}


\section{Introduction}
A uniformly accelerated testing particle in Minkowski vacuum might observe
a background thermal radiation, with the temperature
\begin{equation}
  T_U = {\hbar a\over 2\pi k_B c}\label{Tunruh}
\end{equation}
proportional to the acceleration $a$ of the testing particle. This
remarkable phenomenon, the Unruh effect\cite{dave,unruh}, has been
considered as a connection between quantum($\hbar$),
statistical($k_B$) and relativistic physics. However, since the
quantum correlation functions for free fields are Green's
functions of the corresponding classical field theories, if there
exists thermal characters in a correlation function or a vacuum
expectation value of the stress-energy tensor, corresponding
information should be found in its classical
theory\cite{hacy,mane,hira}. Indeed, in our previous
work\cite{lin}, we explicitly showed a classical correspondence of
Unruh effect in classical relativistic electron theory: the vacuum
expectation value of the energy density for a point-like electron
is identical to its classical self-energy density. In the
classical, point-like framework, the thermal interpretation seems
to be un-necessary and the Unruh temperature $(\ref{Tunruh})$ is
an artifact by identifying the power spectrum to the
Planckian-like ones.

Since the prototype of the detector theory in Rindler space concerns a
point-like monopole detector linearly coupled with scalar
fields\cite{unruh,BD}, we would like to study such kind of models
in the following and try to reach the same conclusion. In Section II we
recall the classical electron theory and the solution of
electromagnetic(EM) field generated by a uniformly accelerated charge. Then
a model of classical monopole detectors for massless scalar fields is given
in Section III for comparing with the literature about the Unruh effect.
Finally we have some discussions in Section IV.

\section{uniformly accelerated electric charge}
To begin with, let us consider the action for the relativistic Lorentz
electron\cite{rohr}
\begin{eqnarray}
  S &=& -m\int d\tau \sqrt{-{dz_\mu\over d\tau}{dz^\mu\over d\tau}}
      -\int d^4 x \sqrt{-g}{1\over 4}F_{\mu\nu}F^{\mu\nu}\nonumber\\
    & & + \int d\tau d^4 x\sqrt{-g} j_\mu (x,\tau) A^\mu(x) ,\label{actEM}
\end{eqnarray}
where $F_{\mu\nu}\equiv D_\mu A_\nu -D_\nu A_\mu$ and the current $j_{\mu}$
is defined by
\begin{equation}
  j_\mu\equiv e{dz_\mu\over d\tau}\delta^4 (x-z(\tau))
\end{equation}
with the coupling constant $e$. The solutions for vector fields can be
written in $A^\mu = A^\mu_{\rm in}+A^\mu_{\rm ret}$ or $A^\mu = A^\mu_{\rm
out}+A^\mu_{\rm adv}$, depending on the choice of boundary conditions. Here
$A^\mu_{\rm ret}$ and $A^\mu_{\rm adv}$ are the retarded and advanced
fields respectively. Let
\begin{eqnarray}
  \bar{A}^\mu &=& {1\over 2}(A^\mu_{\rm in}+ A^\mu_{\rm out})\\
  A^\mu_\pm &=& {1\over 2}(A^\mu_{\rm ret}\pm A^\mu_{\rm adv})
\end{eqnarray}
such that $\bar{A}^\mu =A^\mu_{\rm in} +A^\mu_-$ and $A^\mu = \bar{A}^\mu
+ A^\mu_+ $. It is known that the sigular behavior of the solution are
all present in $A^\mu_+$, hence one re-write
\begin{eqnarray}
  S_{\rm ren}&=&-m\int d\tau\sqrt{-{dz_\mu\over d\tau}{dz^\mu\over d\tau}}
    + \int d\tau d^4 x\sqrt{-g} j_\mu (x,\tau) \bar{A}^\mu(x)\nonumber\\
    & &-\int d^4 x \sqrt{-g}{1\over 4}\left(\bar{F}_{\mu\nu}
    \bar{F}^{\mu\nu}+2\bar{F}_{\mu\nu}F^{\mu\nu}_+ \right)
\end{eqnarray}
by dropping all of the divergent terms. Variation on $z^\mu$, $\bar{A}^
\mu$ and $A_+^\mu$ then gives Lorentz-Dirac equation as well as Maxwell
equations,
\begin{eqnarray}
& & ma^\mu(\tau)=e\bar{F}^{\mu\nu} v_\nu(\tau),\\
& & \partial_\mu \left[\bar{F}^{\mu\nu}(x)+F_+^{\mu\nu}(x)\right]
    =-j^\mu(x),\\
& & \partial_\mu \bar{F}^{\mu\nu}(x)=0,
\end{eqnarray}
where $v^\mu \equiv dz^\mu/d\tau$ is the proper velocity and $a^\mu \equiv
dv^\mu/d\tau$ is the proper acceleration.

An electric charge accelerated in a constant proper acceleration
$|a_\mu a^\mu|^{1/2}=a$ would go along the hyperbolic trajectory
$z'^\mu =(a^{-1} \sinh a\tau,a^{-1}\cosh a\tau,0,0)$,
parameterized by its proper time $\tau$.\footnote{For linearly
accelerated detectors, we use the cylindrical coordinate
$ds^2=-dt^2+dz^2+d\rho^2+\rho^2d\theta^2$ with $c=1$.} The
corresponding EM field was firstly given by Born\cite{born} in
1909, then in 1955 Bondi and Gold\cite{BG} found a more general
solution,
\begin{eqnarray}
  E_z &=& F^{tz}=-{4e\over a^2X^3}\left( a^{-2}+t^2-z^2+\rho^2\right)
    \theta(z+t),\label{bornEz}\\
  E_\rho &=& F^{t\rho} ={8e\rho z\over a^2X^3}\theta(z+t)
    +{2 e\rho\over \rho^2+ a^{-2}}\delta(z+t),\label{bornEr}\\
  B_\phi &=& F^{z\rho} ={8e\rho t\over a^2X^3}\theta(z+t)
    -{2 e\rho\over \rho^2+ a^{-2}}\delta(z+t),\label{bornBp}
\end{eqnarray}
where
\begin{equation}
  X \equiv\sqrt{4a^{-2}\rho^2+\left( a^{-2}+t^2-z^2-\rho^2\right)^2},
  \label{defxi}
\end{equation}
and the step function $\theta (x)$ is defined by
\begin{equation}
  \theta(x)=\left\{\begin{array}{l}1\mbox{ for }x>0\\{1/2}\mbox{ for }
  x=0\\0 \mbox{ for }x<0\end{array}\right. .
\end{equation}
The improvement is that Bondi-Gold solution satisfies Maxwell equations in
the whole spacetime including the event horizon.

In our previous paper\cite{lin}, we noted that the classical energy density
measured by the co-moving observer at $z^\mu = z'^\mu +(0,0,\rho,0)$ can be
related to the Unruh effect, by letting the correlation $\rho = 2 a^{-1}
\sinh(a\Delta/4)$ then performing a Fourier transformation with respect to
$\Delta$. Below we illustrate the same relation in the classical version of
the original detector model for scalar fields.

\section{Unruh-DeWitt type monopole detector}

\subsection{the model}

An Unruh-DeWitt type monopole detector model could be formulated by the
action
\begin{equation}
  S = S_q+S_\phi +S_{\rm int},
\end{equation}
where
\begin{equation}
  S_\phi = -\int d^4 x\sqrt{-g}{1\over 2}\partial_\mu \phi
    \partial^\mu\phi ,
\end{equation}
is the action for a massless scalar field $\phi$,
\begin{equation}
  S_q = \int d\tau\left[{1\over 2}\left(\partial_\tau q\right)^2 -
        V(q)\right],
\end{equation}
is the action for the detector with monopole moment $q$, and
\begin{equation}
  S_{\rm int} = e\int d\tau d^4 x\sqrt{-g} q(\tau )\phi (x)
    \delta^4 (x-x'(\tau)),\label{Sint}
\end{equation}
is the interaction-at-a-point with coupling constant $e$. In this paper we
choose $V = \omega^2 q^2/2$ as a harmonic oscillator for simplicity.
Now the similarity between the Unruh-DeWitt type monopole detector and the 
classical electron theory Eq.$(\ref{actEM})$ is obvious. The main 
difference is that we put the monopole moment here as an extra degree of 
freedom to the motion, while the dipole moment in electron theory 
$(\ref{actEM})$ is directly related to the position of the electron. Hence 
the trajectory of the monopole detector can be arbitrary in our model.

To obtain a renormalized action, one can decompose $\phi$ in the same
fashion of the classical electron theory, namely,
\begin{equation}
  \phi = \bar{\phi}+\phi_+ ,
\end{equation}
where $\bar{\phi}$ and $\phi_+$ are homogeneous and inhomogeneous
solutions of $\phi$ respectively.
Then, omitting the singular terms, the regular part reads
\begin{eqnarray}
  S_{\rm ren}&=& \int d\tau\left[ {1\over 2}(\partial_\tau q)^2-
     V(q)\right]\nonumber\\
& &-\int d^4 x\sqrt{-g}{1\over 2}\left[ \partial_\mu\bar{\phi}\partial^\mu
   \bar{\phi} +2\partial_\mu\phi_+\partial^\mu\bar{\phi}\right]\nonumber\\
& &+e\int d\tau d^4 x\sqrt{-g}q(\tau)\bar{\phi}(x)\delta^4(x-x'(\tau)),
\label{Sren}
\end{eqnarray}
which yields the equation of motion and field equations as follows,
\begin{eqnarray}
  & &\partial_\tau^2 q + V'(q) = e\bar{\phi}(x'(\tau)),\label{eqq}\\
  & &\Box ( \bar{\phi}+\phi_+ ) = e\int d\tau q(\tau)
    \delta^4\left(x-x'(\tau)\right),\label{eqphi}\\
  & &\Box \bar{\phi} = 0,
\end{eqnarray}
where $V'(q)\equiv \delta V/\delta q =\omega^2 q$. The first equation is
equivalent to the equation of motion for a driven oscillator. The second
equation descibes the field configuration generated by a point source with
a time-varying charge. The third is the homogeneous equation for the scalar
field.

\subsection{static case in Minkowski space}

Suppose the detector locates at the origin, namely, $x'(\tau)=(t,0,0,0)$.
In Minkowski space, the static solution requires that $\tau = t$, $q(\tau)=
q_0$ and $\bar{\phi}(x'(\tau))=\phi_0$, where $q_0$ and $\phi_0$ are
constants of $t$. Then Eq.$(\ref{eqq})$ gives $\omega^2 q_0=e\phi_0$
and Eq.$(\ref{eqphi})$ gives
\begin{equation}
  \phi_+ (x) = e q_0 \int d\tau G_+ (x-x'(\tau))
  = {eq_0\over 8\pi r}\label{phi0+}
\end{equation}
such that
\begin{equation}
  \phi = \left. {e^2\phi_0\over 8\pi\omega^2 r}
  +\bar{\phi}\right|_{\bar{\phi}(x'(\tau))=\phi_0}\label{phi00}
\end{equation}
where $r\equiv\sqrt{\rho^2+z^2}$ is the radius in spherical coordinate and
$G_+ (x-x')\equiv (4\pi)^{-1}\theta (t-t'(\tau))\delta(|x-x'(\tau)|^2)$ is
the retarded Green's function of $\phi_+$. Since the solution above has the
time-reversal symmetry, the retarded field starts at $t=-\Delta/2$
is equal to the advanced field ends at $t=+\Delta/2$. Then one can
introduce the correlation $r=|\pm\Delta/2|$ and write
\begin{equation}
  \phi_+(-\Delta/2)\phi_+(\Delta/2) = {e^2q_0^2\over 16\pi^2 \Delta^2}.
  \label{phi+-cl}
\end{equation}
On the other hand, given the Hadamard's elementary function for massless
scalar field, $D^{(1)}(x-x'(\tau))=\left< 0|\{\phi(x)\phi(x')\}|0\right>
=\hbar/2\pi^2(x^\mu-x'^\mu(\tau))(x_\mu-x'_\mu(\tau))$\cite{BD}, one has
\begin{equation}
  \left< 0|\{\phi(t=-\Delta/2)\phi(t'=\Delta/2)\}|0\right> =
  -{\hbar \over 2\pi^2 \Delta^2},
\end{equation}
whose $\Delta$-dependence is the same as Eq.$(\ref{phi+-cl})$.

One may further compare physical quantities such as energy densities from
both sides. The modified stress-energy tensor in Minkowski space is defined
by\cite{BD}
\begin{eqnarray}
  T_{\mu\nu}&=&\left( 1-2\xi\right)\phi_{,\mu}\phi_{,\nu}-2\xi\phi
  \phi_{;\mu\nu}\nonumber\\ & &+ \left( 2\xi
  -{1\over 2}\right)g_{\mu\nu}g^{\rho\sigma}\phi_{,\rho}\phi_{,\sigma}
  +{\xi\over 2}g_{\mu\nu}\phi\Box\phi ,
\end{eqnarray}
where $\xi$ is a constant parameter. Here $T_{tt}$ with different values
of $\xi$ are different by total derivatives of space, hence their local 
physics are the same\cite{full}. Now a straightforward calculation gives 
the classical energy density
\begin{equation}
  T_{tt}(t,r)={e^2 q_0^2 (1-4\xi)\over 128\pi^2 r^4}=
  {e^2 q_0^2(1-4\xi)\over8\pi^2\Delta^4}, \label{Ecl0}
\end{equation}
at the point $(t,r,0,0)$ in spherical coordinate,
while the expectation value of the energy density of the detector reads
\begin{eqnarray}
  \left<T_{tt}\right> &=& {\alpha'\over 2}\lim_{x^\mu\to x'^\mu}\left[
  {1\over 2}\partial_t\partial_{t'}-\left(2\xi-{1\over 2}\right)
  (\partial_x\partial_{x'}+\partial_y\partial_{y'}+\partial_z\partial_{z'})
  \right.\nonumber\\ & & \left.
  -{3\xi\over 2}\partial_t^2-{\xi\over 2}(\partial_x^2+
   \partial_y^2+\partial_z^2)\right] D^{(1)}(x,x')\nonumber\\
  &=& \lim_{\Delta\to 0}{3\alpha'\hbar\over 2\pi^2\Delta^4}.\label{Eq0}
\end{eqnarray}
As suggested by Mane\cite{mane}, the vacuum power flux for the EM field 
quantized in rotating frame\cite{hacy} can be identified to the classical 
synchrotron radiation from a rotating electron, by multiplying an overall 
factor proportional to the EM fine-structure constant $\alpha=e^2/\hbar c$ 
to the former. This factor, originated from the coupling between electrons 
and photons, is natural in quantum electrodynamics, but not so obvious in 
the semi-classical EM theory with point-like sources. The same 
identification is confirmed by Hirayama and Hara\cite{hira} and works well 
in our previous article\cite{lin}. Hence we recognize that above quantum 
expectation value is actually to the classical order. To identify both 
results, we simply compare Eq.$(\ref{Eq0})$ with Eq.$(\ref{Ecl0})$ and 
choose the overall factor $\alpha'=e^2 q_0^2(1-4\xi)/12\hbar$.

\subsection{acceleration and Planck factor}

Next let us consider the detector moving in a uniform acceleration $a$.
Again the trajectory is $x'(\tau)=(a^{-1}\sinh a\tau, a^{-1}\cosh a\tau,0,
0)$. Suppose the detector is in equilibrium with the background, which 
corresponds the classical solutions static with respect to $\tau$. 
Requiring constant solutions that $q(\tau )=q_0$ and $\bar{\phi}
(x'(\tau))=\phi_0$ along the trajectory, where $q_0$ and $\phi_0$ are 
constants of $\tau$ subject to $\omega^2 q_0=e\phi_0$ from 
Eq.$(\ref{eqq})$, the total solution of $\phi$ then reads
\begin{equation}
  \phi = \left. {e^2\phi_0\over 4\pi\omega^2 a X}\theta(z+t)
  +\bar{\phi}\right|_{\bar{\phi}(x'(\tau))=\phi_0}\label{phicl}
\end{equation}
where $X$ has the same definition as Eq.$(\ref{defxi})$.

In the region with $z+t>0$ and $z-t>0$, $(\ref{phicl})$ has the
time-reversal symmetry, which makes the retarded field equal to the
advanced field like the static case in Minkowski space (clearly this
property is true only when the acceleration is uniform). At $z^\mu = (0,a^{
-1},\rho,0)$, the inhomogeneous part of $\phi$ has
\begin{equation}
  \phi_+(-\Delta/2)\phi_+(\Delta/2) =\left({e q_0\over 4\pi a}\right)^2
  {1\over\rho^2(\rho^2+4a^{-2})}={e^2q_0^2a^2\over 16\pi^2\sinh^2
  (a\Delta /2)}
\end{equation}
if one substitute the correlation $\rho=2a^{-1}\sinh(a\Delta/4)$ given in
Ref.\cite{lin}. This is identical to the correlation function for this
case in terms of Hadamard's elementary function up to an overall factor,
hence acquires the same thermal character.

Recall that the first order perturbation theory of a quantum detector with
the same interacting action $S_{\rm int}$ (un-renormalizd, Eq.$(\ref{Sint}
)$) has the transition probability per unit proper time\cite{unruh,BD},
\begin{eqnarray}
P&=&{e^2\over 2\pi}\sum_{E\not=E_0}\left|\left< E|q(0)|E_0\right>
  \right|^2\int^\infty_{-\infty}d(\Delta\tau) e^{-i(E-E_0)(\Delta\tau)}
  D^{(1)}(\Delta\tau)\nonumber\\
&=& {e^2\over 2\pi}\sum_{E\not= E_0} {(E-E_0)\left|\left< E|q(0)
  |E_0\right>\right|^2\over e^{2\pi(E-E_0)/a}-1}+{\rm singular\ terms},
\end{eqnarray}
if the detector was prepared in its ground state at past null infinity
$(\tau\to -\infty)$.
It is clear that the Planck factor
in $P$ is totally from the regular part of the correlation function
$D^{(1)}\sim 1/\sinh^2 (a(\Delta\tau) /2)$ in $\Delta\tau\to 0$ limit.
Nevertheless, in our classical theory, no transition rate can be defined 
for the detector at all. To see the thermal character of the classical 
uniformly accelerated detector, one has to look into the energy density.
At $z^\mu =(0,a^{-1},\rho,0)$, the energy density for the inhomogeneous 
part of $\phi$ reads
\begin{equation}
  T_{tt}(\phi_+)={e^2 q_0^2\left[1-4\xi+(1-6\xi)a^2\rho^2\right]
  \over 8\pi^2\rho^4(4+a^2\rho^2)^2},\label{Ecl}
\end{equation}
while the quantum expectation value of $T_{tt}$ is
\begin{eqnarray}
  \left< T_{tt}\right> &=&{\alpha'\hbar a^4\over32\pi^2}\lim_{\tau_\pm
  \equiv\pm\Delta/2\to 0}{1+2\cosh a(\tau_+ +\tau_-)\over\sinh^4 [a(\tau_+
  -\tau_-)/2]}\nonumber\\
  &=& \lim_{\Delta\to 0}{e^2 q_0^2 a^4(1-4\xi )\over 128\pi^2\sinh^4
    (a\Delta /2)}.\label{EQ}
\end{eqnarray}
After one substitutes $\rho=2 a^{-1}\sinh(a\Delta/4)$ into Eq.$(\ref{Ecl})
$, above two energy densities are exactly the same when $\xi = 1/6$,
which corresponds to a conformal invariant scalar field theory.
Furthermore, the $\Delta$-dependence in Eq.$(\ref{EQ})$ is identical to
those for classical uniformly accelerated electric charges (see
Eq.(13)-(16) in Ref.\cite{lin}), hence one can apply the same approach in
renormalization and find the ``vacuum energy" being $11e^2 q_0^2a^4/17280
\pi^2$ for $\xi=1/6$. Again the Planck factor can be obtained, for each
value of $\xi$, by performing a Fourier transformation
with respect to $\Delta$ on the classical energy density $(\ref{Ecl})$.

\section{Discussion}

\subsection{existence of specific motion}

For our Unruh-DeWitt type monopole detector theory, the accelerating motion
of the detector is actually arbitrary. Nevertheless, the same arbitrariness
is not true in detector theories for EM or gravitational fields. In the EM
case, the dipole moment coupled to the homogeneous EM field $\bar{A}^\mu$
is directly related to the position of the charge, rather than a degree of
freedom independent of motion in our monopole model. Whether there exists a
specific accelerating motion depends on the existence of this solution of
the Lorentz-Dirac equation together with Maxwell equations. Stationary
circular motion around the nucleus, for example, is impossible for a
classical electric charge\cite{jack}, hence the vacuum state strictly
defined by the corresponding coordinate system has no physical meaning for
a point-like electric charge.

But one still has the right to consider the instantaneous ``vacuum
stress-energy" for any accelerated detector, as we pointed out in our
previous article\cite{lin}. The reason is that the instantaneous ``vacuum
stress-energy" can be locally defined by the co-moving observer, while the
power spectrum needs the knowledge about the whole history to perform the
Fourier transformation.

\subsection{particle detector creates particle}

There might exist more than one timelike Killing vectors in some 
spacetimes. Along these time axes, one can formulate different Hamiltonian 
theories, whose vacua might disagree with each other. In a pure field 
theory consideration, there is no guideline to determine which vacuum is 
correct, unless one imposes a detector or a testing particle, fixes 
physical boundary conditions, then observes the response of the detector.

So far what one can do in testing the detector theory is also to accelerate 
some elementary particles without structure, such as electrons, then 
measure their responses in laboratories\cite{bell}. Therefore only the back 
reaction of the detector is interesting and measurable for experimental
physicists, yet whether the detector really see a thermal bath is not 
important here.

When the back reaction of the detector on the field configuration is
taken into account, an classical observer co-moving with the uniformly
accelerated detector may conclude that the detector experiences a thermal
bath of Rindler particles rather than a zero-temperature vacuum, by
measuring the static field strength around the detector then introducing 
the correlation between their distance ($\rho$) and the clock of the 
detector ($\tau$ or $\Delta$). Hence the particle population that the 
detector experiences, speculated by the co-moving observer, depends on how 
the detector (and the co-moving observer) moves. In this sense the particle 
detector itself is the creator of the particle.

\subsection{Can a classical point-like monopole detector detect its
acceleration?}

No matter how close the co-moving observer is to the point-like detector,
they would never overlap by definition. While the self-energy density for
the scalar field is singular right at the position of the detector, the
internal state of the classical monopole detector is actually finite and
static with respect to its proper time. The constant solution for monopole
moment $q=q_0$ corresponds to the lowest ``effective potential energy" $V-
{\cal L}_{\rm int} = \omega^2 q_0^2/2-e\phi_0 q_0$ for the detector. Since
it does not exert work, it seems that one can relate this ``effective
potential energy" to the entropy. However, $q_0$ and $\phi_0$ are
parameters independent of the proper acceleration $a$. The $q_0$ or $\phi_0
$ in Minkowski case Eq.$(\ref{phi00})$ for $a=0$ can be the same as the
one in Rindler case Eq.$(\ref{phicl})$ for $a\not=0$, if the $\bar{\phi}
(x'(\tau))=\phi_0$ are assigned the same value in both cases. This means
that a classical {\it point particle} has no idea about its acceleration
from its monopole moment. To ``know" the acceleration classically, an
extended object with a finite size or a point-like detector with an
outer observer co-moving at a non-zero distance is necessary. In the latter
case the Planckian-like spectrum measured by the co-moving observer is
simply a signal that the detector-observer pair is uniformly accelerated,
just like the water-level of an accelerating bucket. Indeed, an
accelerating bucket would find its water-level is oblique due to the tidal
force, but if the bucket shrinks into a point, it cannot find any evidence 
about the acceleration by itself.

Even if the uniformly accelerated detector does not see the thermal bath
according to its monopole moment, there still exist a Planck factor in the
power-spectrum of the ``vacuum energy", speculated by the co-moving 
observer with a particular choice of variable. This again suggests that the 
thermal interpretation may not be necessary for classical point-like 
detectors.

\begin{acknowledgments}
I wish to thank T. Hirayama for many helpful comments. I would also like 
to thank J. Nester, C. I Kuo, W.-L. Lee, J.-P. Wang for discussions.
\end{acknowledgments}

\end{document}